\documentclass[pre,twocolumn,aps,superscriptaddress,longbibliography]{revtex4-1}
\usepackage{amsmath}
\usepackage{amssymb}
\usepackage{graphicx}
\usepackage[usenames,dvipsnames]{color}
\usepackage{bm}
\usepackage{times}
\usepackage{hyperref}
\usepackage{float}
\usepackage{verbatim}
\usepackage{lipsum}
\usepackage{braket}
\usepackage{soul}
\usepackage[T1]{fontenc}
\usepackage[latin1]{inputenc}
\usepackage[table]{xcolor}
\hypersetup{
  colorlinks=true,
  citecolor=blue,
  linkcolor=blue,
  urlcolor=blue}

\definecolor{orange(colorwheel)}{rgb}{1.0, 0.5, 0.0}
\definecolor{maroon(html/css)}{rgb}{0.5, 0.0, 0.0}
\definecolor{lightgray}{gray}{0.9}

\begin{document}

\title{Fine-grained domain counting and percolation analysis in 2D 
       lattice systems with linked-lists} 

\author{Hrushikesh Sable}
\affiliation{Physical Research Laboratory,
             Ahmedabad - 380009, Gujarat,
             India}
\affiliation{Indian Institute of Technology Gandhinagar,
             Palaj, Gandhinagar - 382355, Gujarat,
             India}
\affiliation{Department of Physics, Virginia Tech, 
	     Blacksburg, Virginia 24061, USA}
\author{Deepak Gaur}
\affiliation{Physical Research Laboratory,
             Ahmedabad - 380009, Gujarat,
             India}
\affiliation{Indian Institute of Technology Gandhinagar,
             Palaj, Gandhinagar - 382355, Gujarat,
             India}
\author{D. Angom}
\affiliation{Physical Research Laboratory,
             Ahmedabad - 380009, Gujarat,
             India}
\affiliation{Department of Physics, Manipur University, Canchipur - 
             795003, Manipur, India}
\date{\today}

\begin{abstract}

 We present a fine-grained approach to identify clusters and perform
percolation analysis in a 2D lattice system. In our approach, we develop
an algorithm based on the linked-list data structure whereby the members of
a cluster are nodes of a path. This path is mapped to a linked-list.
This approach facilitates unique cluster labeling in a lattice with a
single scan. We use the algorithm to determine the critical exponent
in the quench dynamics from the Mott insulator to the superfluid phase of bosons
in 2D square optical lattices. The results obtained are consistent with the
Kibble-Zurek mechanism. We also employ the algorithm to compute the correlation
length using definitions based on percolation theory and use it to
identify the quantum critical point of the Bose Glass to superfluid
transition in the disordered 2D square optical lattices. In 
addition, we compute the critical exponent $\nu$ which quantify the 
divergence of the correlation length $\xi$ across the phase transition and the 
fractal dimension of the hulls of the superfluid clusters.
\end{abstract}
\maketitle


\section{Introduction}
\label{intro}
 There are strong motivations across research disciplines to develop novel 
approaches and computational methods to study the percolation theory. 
The percolation theory provides a simple and unifying framework to understand 
clustering of particles in a medium. It has wide applications like the 
permeation of fluid in porous media \cite{Broadbent_57}, spontaneous 
magnetization of dilute ferromagnets \cite{elliott_60}, and polymer 
gels \cite{flory_53}, to mention a few. The study of the clustering or 
connectivity of the particles gains importance as it determines the macroscopic 
properties of the system. For example, the electrical conduction through a 
composite mixture of conducting and insulating materials is described by 
the percolation analysis of the conducting material in the mixture.

 A generic percolation problem consists of an infinite lattice populated with
two classes of lattice sites at random. These are denoted as occupied and 
unoccupied sites \cite{stauffer_92, sahimi_21}. A group of occupied sites
connected through bonds forms a cluster. The probability of a site being
occupied determines the distribution and size of the clusters. The system 
undergoes a percolation phase transition when the probability exceeds a 
critical value. Then, there exists a spanning cluster which extends from one 
edge of the system to the opposite edge. Although the percolation problem as
stated is straightforward, analytical approaches are limited. It is, however,
possible to gain an understanding of the system using numerical methods. A 
prominent algorithm used in percolation analysis is the Hoshen-Kopelman (HK) 
cluster multiple labeling algorithm \cite{hoshen_76}. 
The application of the percolation analysis and of the HK algorithm are in
diverse fields like - food and chemical 
engineering \cite{moreira_96, zhang_96}, ecology \cite{berry_94} and biology 
\cite{eddi_96}. The basic essence of the algorithm is to scan through the 
lattice and identify the occupied sites. During the scan, the occupancy of 
neighboring sites are also checked and the connected occupied sites are 
assigned a cluster label. The key point of the algorithm is that the 
neighboring sites are allowed to have different labels, but with a record 
that they belong to the same cluster. There are now several variations of 
the HK \cite{hoshen_97, frijters_15, constantin_97, moloney_03, teuler_2000}
including a proposal to use linked-list to 
group the clusters belonging to the same domain \cite{sendra_06}.
 
 In this work, we present a fine-grained algorithm of cluster labeling and 
describe its application in the percolation analysis of two-dimensional (2D) 
lattices. The algorithm employs the linked-list data structure and we refer to 
it as domain counting (DC) algorithm. Using which we can define
a path connecting all the sites belonging to a cluster. This facilitates the 
scanning of the lattice in a single scan. The path constructed 
is fine-grained as it links the sites . Such an approach facilitates the 
analysis of cluster properties. This is in contrast to the HK algorithm, where 
the equivalence class built links different cluster labels of the same 
underlying cluster, and is a coarse-grained linking. Different studies
that can be done with this algorithm include determination of boundaries, easy 
identification of spanning
cluster, and calculation of various cluster properties like center of mass,
radius of gyration, and correlation length. Our approach is well suited
for detailed analysis of results encountered in the studies of optical 
lattices where we obtain a set of configurations and wish to examine it 
using the tools from percolation theory. However, it must be emphasized that
to simulate percolation the Newman-Ziff algorithm \cite{newman_00} is  
the method of choice. 

We address two important problems in the physics of ultracold bosonic atoms in
optical lattices using the algorithm. First, we study the quantum quench 
dynamics of  bosons in the optical lattices from the Mott insulator (MI) to 
the superfluid (SF) phase employing the time-dependent Gutzwiller mean-field 
theory. Using our method we identify the clusters or domains and show that 
the number of domains follows a power-law dependence on the quench rate. This 
is consistent with  the Kibble-Zurek 
mechanism (KZM) \cite{kibble_76,kibble_80, zurek_85, zurek_96, delCampo_14}. 
It is to be mentioned that recent works on ultracold atoms have 
reported similar results in quantum quenches across different quantum 
phases \cite{shimizu_misf_18, shimizu_dwss_18, shimizu_dwsf_18, zhou_20}. 
Further, as to be expected, we show that the dependence of the defect density 
on the quench rate has the same power-law exponent as the number of domains. 
The second study pertains to the critical properties of the Bose glass (BG) to
SF transition of bosons in disordered optical lattices. The BG phase is 
insulating yet compressible and it is characterized by the SF {\em puddles} 
with an MI background \cite{fisher_89, buonsante_07, pollet_09, pal_19}. 
Using our method we compute the geometrical properties of the SF clusters.
As mentioned, our fine-grained method stores locations of all the sites 
of a cluster. From this information, we calculate the percolation correlation 
length of the system across the BG-to-SF transition. As expected, the 
correlation length peaks at the transition point and has power-law 
dependence on the reduced hopping strength. The critical exponent $\nu$ 
which quantifies the power law reveals that the BG-to-SF transition belongs 
to the universality class of 2D random percolation. We also 
compute the fractal dimension of the hulls of SF clusters near the 
transition point.

The remainder of this article is organized as follows. We first introduce 
linked-lists and discuss mapping of the domains to a linked-list
in Sec.~\ref{linkedlist_th}. We then discuss the algorithm of our method
in the Section \ref{algo_theory}. We describe the identification of the 
boundary of the cluster in Sec.~\ref{bndry}. In Sec.~\ref{comparison_section}, 
we discuss the comparison of the presented algorithm
with other standard algorithms in the percolation theory. 
In Sec.~\ref{application}, 
we discuss the application of our method to the MI-SF quench dynamics.
Then, in the Sec.~\ref{percol_bg}, we present the study of the BG-to-SF 
transition from the perspective of percolation theory. We summarize our 
main results and present discussions in Sec.~\ref{conclusion}.


\section{Mapping domain to a linked-list}
\label{linkedlist_th}

 For a better description of our method, we consider a 2D square lattice and 
each of the lattice sites are labeled at random with either $0$ or $-1$. 
In the percolation theory, the site labeled $0$ ($-1$) corresponds to 
occupied (unoccupied) site. Accordingly, we may assume that the label 
$0$  ($-1$) occurs with the probability $p$ ($1-p$). 
As mentioned earlier, the task at hand is to identify and enumerate the domains 
with either of the two labels. For illustration let us take the domains of 
sites with the label $0$. 
\begin{figure}[ht]
  \includegraphics[width=9cm]{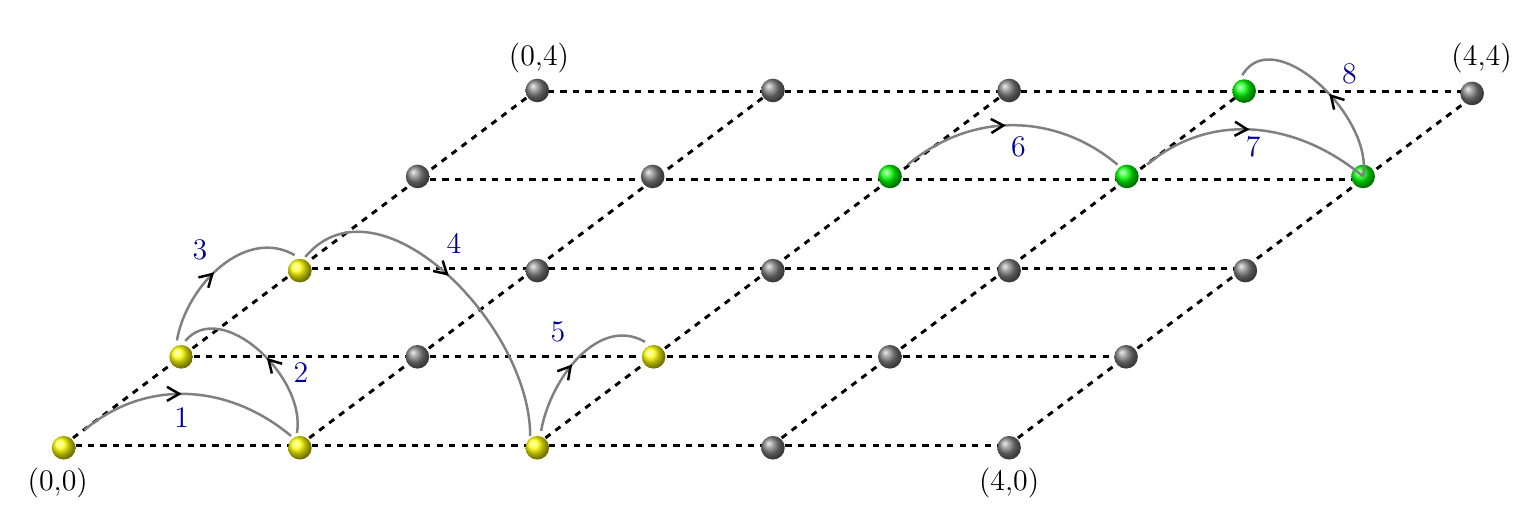}
  \caption{A schematic diagram to illustrate the paths defining a cluster in 
	a 2D lattice. The arrows indicate the links between sites constituting 
	a domain.
	Here distinct domains are identified with different colors which are
	representative of domain labels, while the gray-shaded lattice sites 
	are of label $-1$. The numbers over the arrows represent the 
	chronological sequence in linking the sites.
	}
  \label{xyloc}
\end{figure}
The lattice sites are denoted by $(i, j)$, where $i$ is the column index and 
$j$ is the row index. These correspond to the $x$- and $y$-axis coordinates, 
respectively, in the $xy$-plane. The central idea of our approach is to use 
the concept of linked-list data structure \cite{newell_57} to describe a 
domain. It stores a data sequence in noncontiguous memory locations. 
Each element of the sequence is stored in
a node of the list and each node has a reference or a link to the memory 
location of the next element in the sequence. This continues until the last 
element of the sequence. The advantage of using linked-list is the ease of 
updating it to insert a new node or  merging multiple linked-lists. In the 
present work, to define a domain, each lattice site of the system is uniquely 
mapped to the node of a unidirectional linked-list. The  node is then
linked to another lattice site which belongs to the same domain. This way 
each domain is represented by one linked-list. Thus, once we know the 
starting node of a linked-list, we can traverse through all the lattice sites 
belonging to a domain. For this reason, the linked-list associated with a 
domain can also be thought of as a {\em path} traversing through it. Here 
after, we use path to refer to the linked-list. Such a path passes through 
each lattice site in the domain once. Our main objective is, 
then, to enumerate the number of such distinct paths corresponding to the 
domains of lattice sites with the label $0$. Then the number of such paths 
is the number of domains.
 
To define the link in the node corresponding to the $(i, j)$ lattice site, we 
introduce two  variables $x_{i, j}$ and $y_{i, j}$. The variable $x_{i, j}$ 
($y_{i, j}$) is the location along the $x$ ($y$) direction of the next lattice 
site in the path. These variables should have well-defined values for the 
lattice sites at the beginning and intermediate nodes of a path. But this is 
not required for the last node of a path. To distinguish the end node we assign 
negative integers to these two variables for this node. Since we use 
unidirectional linked-list it is essential to define the location of the first 
lattice site in a path or the first node of the path. For this we introduce two 
variables for each path $\alpha^k$ and $\beta^k$; these are the $x$ and $y$ 
locations of the first lattice site of the $k$th path, respectively. To 
facilitate further analysis of the domains, we denote the total number of 
nodes or lattices sites in the $k$th domain by ${\cal N}^k$. 
The two variables 
$\gamma^k$ and $\delta^k$ are introduced to define the $x$ and 
$y$ location of the site corresponding to the last node of the path. Suppose 
$\{x^k_{i, j}, y^k_{i, j}\}$ is the set of the links of the $k$th path, then 
the set of variables 
\begin{equation}
 \{\alpha^k, \beta^k, x^k_{i, j}, y^k_{i, j}, \gamma^k, \delta^k, {\cal N}^k\},
\end{equation} 
defines the $k$th path or the domain uniquely. The schematic diagram of 
typical paths defining clusters on a 2D square lattice are shown  in
Fig. \ref{xyloc}.


\section{Domain identification}
\label{algo_theory}

  To identify the domains, we check the label at each lattice site
column wise. We start the scan from the left edge. That is, the column scanning
is done left to right or in an increasing order of the lattice site index 
along the $x$ direction. During the scan, we check the label at a 
lattice site, say $(i, j)$. 
If the label of the site is $0$, then the variable $k$  is
incremented by $1$, and the site is relabeled as $k$.
This identifies the lattice site as a member of the $k$th 
domain and the site is linked to the path. This number, following earlier 
description, is also the sequence number of the domain. There is no change in 
the label if the lattice site is already identified as the member of a domain. 
Then, we scan the label of the right nearest-neighbor lattice site 
$(i+1, j)$. If the label of this site is also $0$, then it is relabeled as 
$k$. This identifies the lattice site  $(i+1, j)$ as a member of the same 
domain. Accordingly, we update the path to include the site. Then the scan is 
continued to the lattice site $(i, j+1)$, the nearest neighbor above 
$(i, j)$. In case the label of the lattice site $(i, j)$ is $-1$, then the 
scan proceeds to $(i, j+1)$. That is, the label of the right nearest neighbor 
is not checked.  This process is repeated until the topmost lattice site of the 
column is reached. Then, we move to the next column and  continue this 
process until the entire system is covered. Two distinct cases arise. First, 
the left edge column, the column scanned first. Since, this is the starting 
column, domains are yet to be identified. 
The second case concerns the columns in the bulk and the right edge column. 
To record the number of domains identified we use the counter $\kappa$.
\begin{figure}[ht]
  \includegraphics[scale = 0.5]{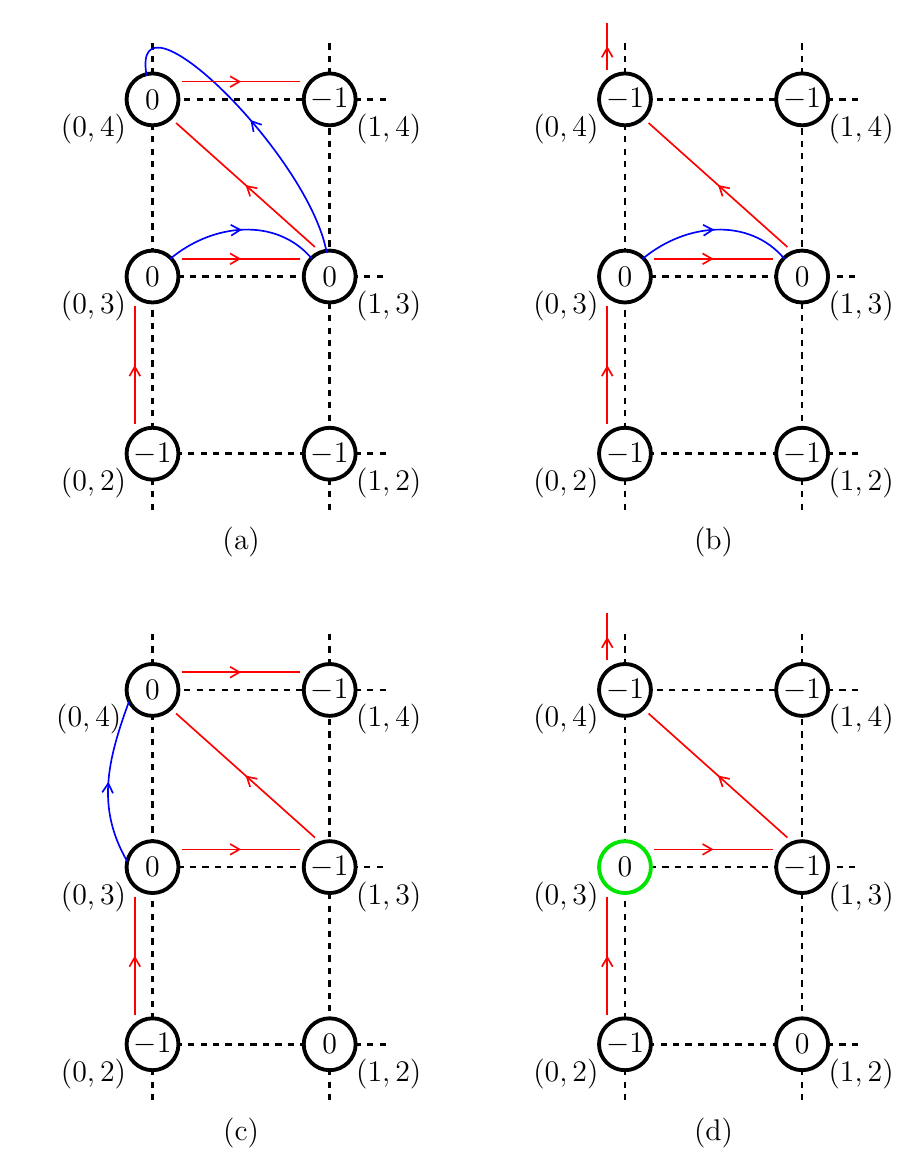}
  \caption{A schematic diagram to illustrate of the possibilities
        arising in
       the scan along the lattice sites of the leftmost column. The red arrows
       indicate the scanning direction, and the blue arrows represent the
       intersite links. The sites are represented by circles and the labels
       are shown within the circles. Panels (a)-(c) represent different cases
       of linking the site $(0,3)$ with its right and/or upper neighbors.
       Panel (d) represents the possibility of a domain constituting a 
       single site.}
  \label{left_ed}
\end{figure}
\subsection{Left edge column}
\label{left_edge_col}

 At the beginning of the scan, we initialize the variable $\kappa$ to zero. 
It is incremented by one when a new domain is encountered. For example,
assume that the lattice site $(0,3)$ is the first lattice site along the 
column which has label $0$. During the scan, on reaching this lattice site, 
we increment $\kappa$ to $1$ and relabel the lattice site $(0,3)$ as $1$. 
This is a newly identified domain; hence, it is the first and the last 
node of the path for the domain with $k=1$. Accordingly, set the first 
node variables $\alpha^1=0$ and $\beta^1=3$ and the last node 
variables  $\gamma^1=0$ and $\delta^1=3$. In addition, the counter for the 
number of the nodes in the domain is updated as ${\cal N}^1=1$. Since this is 
the last node of a path we set $x^1_{0,3}=-2$ and $y^1_{0,3}=-2$. The choice 
of $-2$ is arbitrary. It is a number which is not assigned to any of the 
variables. Then, we check the label of the right nearest-neighbor lattice 
site $(1,3)$. There are four possible outcomes as follows:
{\em Case A}: If this lattice site has $0$ label, then it is relabeled as 
$1$. The end node of the path is shifted to $(1,3)$ by assigning 
$x^1_{1,3}=x^1_{0,3}$ and $y^1_{1,3}=y^1_{0,3}$. Accordingly, update the last 
node variables to $\gamma^1=1$ and $\delta^1=3$. The path is then updated by 
linking the lattice site $(0,3)$ to $(1,3)$ by setting $x^1_{0,3}=1$ and 
$y^1_{0,3}=3$. Then we continue the scan along the column to the lattice 
site $(0,4)$ located above $(0,3)$. If this too is $0$, then we extend the path 
to this lattice site by making this the end node with the assignment 
$x^1_{0,4}=x^1_{1,3}$ and $y^1_{0,4}=y^1_{1,3}$ and update the end node 
variables to $\gamma^1=0$ and $\delta^1=4$. Then connect it to the lattice 
site $(1,3)$ by redefining $x^1_{1,3}=0$ and $y^1_{1,3}=4$. This possibility 
is schematically shown in Fig.~\ref{left_ed}(a). \\
{\em Case B}: The lattice site $(1,3)$  is labeled zero but not $(0,4)$. 
This possibility is depicted schematically in Fig.~\ref{left_ed}(b). Then, the
last step of variable assignments in Case A is not required.
{\em Case C}: The lattice site $(1,3)$ is labeled $-1$, but $(0,4)$ is 
labeled $0$. This is similar to the Case A, but without the intermediate
step of linking the lattice site $(1,3)$. The situation is schematically 
shown in Fig.~\ref{left_ed}(c). \\
{\em Case D}: This is the last case and corresponds to the situation when 
both the lattice sites $(1,3)$ and $(0,4)$ are $-1$. Then the 
$(0,3)$ is an isolated domain as shown in Fig.~\ref{left_ed}(d).
\begin{figure}[ht]
  \hspace*{-0.9cm}
  \includegraphics[scale=0.5]{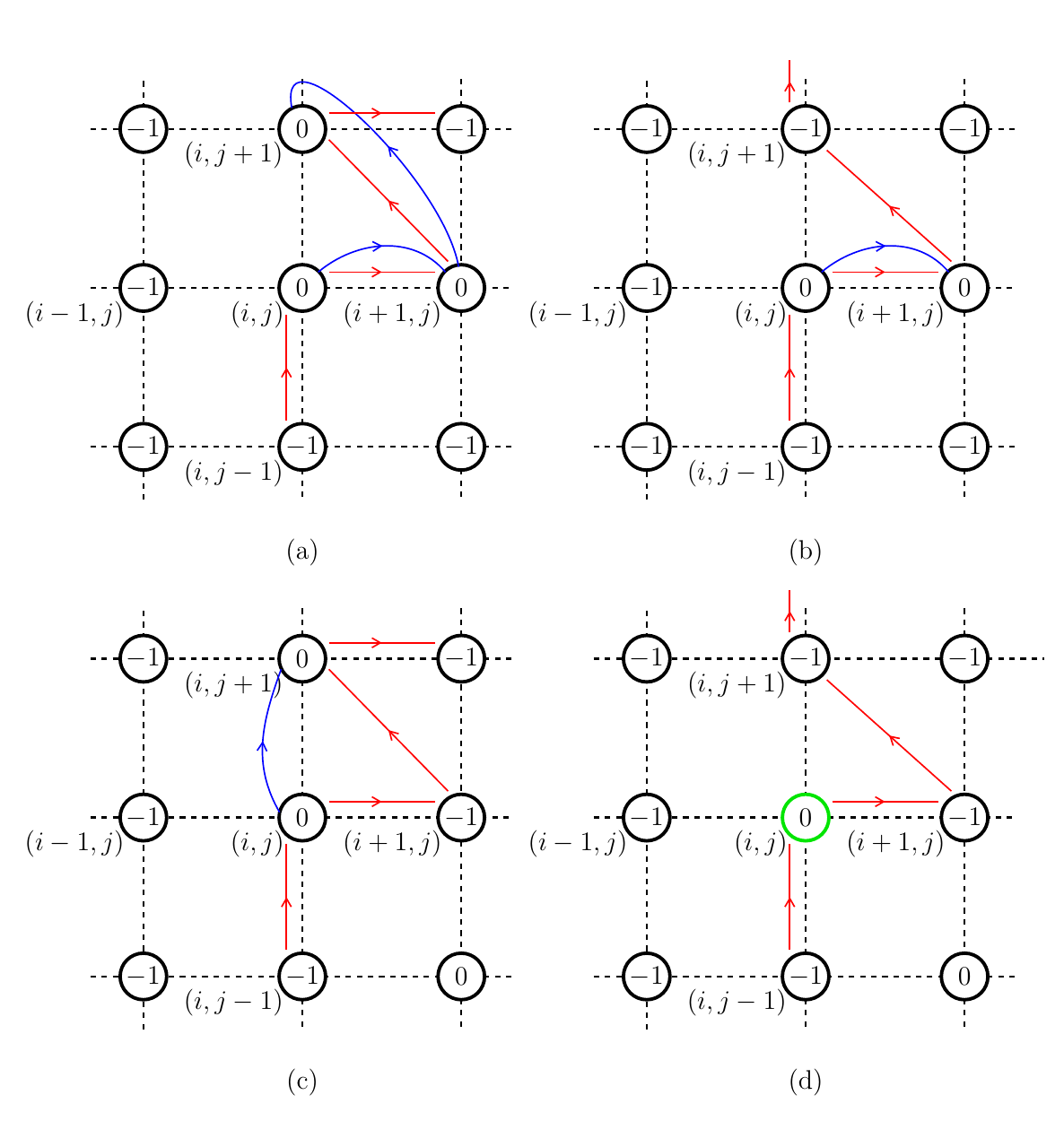}
  \caption{A schematic diagram to illustrate the formation of 
	   a new domain comprising of sites in the bulk. The site ($i,j)$, 
	   labeled $0$, is the starting node of the path. The steps of 
	   scanning and linking the sites in the path are similar to that in 
	   Fig.~\ref{left_ed}}.
  \label{new_dom}
\end{figure}
\subsection{Column in the bulk and right edge}
\label{bulk_right}

 The general steps of scanning the remaining columns of the system are the 
same. For illustration, as a general case, let us consider the scanning of the 
$i$th column and assume that we have reached the $j$th row in the column. That
is, the lattice site to be scanned is $(i,j)$. If the label of the lattice 
site is $-1$, then this is a trivial case and the site does not belong to any 
domain. The scan can continue to the next lattice site $(i,j+1)$ in the 
column. In case the label of the lattice site $(i,j)$ is not $-1$, then, 
there are three possible outcomes of the scan.
\subsubsection{New domain}
\label{new_domain}
Consider that the label of the lattice site $(i,j)$ is $0$, then the current 
value of $\kappa$ is incremented by one. The new value of $\kappa$ is taken as 
the value of $k$, and the lattice site $(i,j)$ is relabeled with this value.
Like in the case of the left edge, set $x^k_{i,j}=-2$ and $y^k_{i,j}=-2$ for 
the path and ${\cal N}^k=1$ for the number of members. 
The first (last) node variables of the path are set as 
$\alpha^k=i$ ($\gamma^k=i$) and $\beta^k=j$ ($\delta^k=j$).
We, then, scan the lattice site $(i+1,j)$, and 
followed by $(i, j+1)$. Similar to the case of the left edge column, discussed 
earlier, we can have four possible outcomes. The only difference is, in each 
of the cases the lattice site $(0,3)$ is replaced by $(i,j)$. The four cases 
are schematically shown in Fig.~\ref{new_dom}. 

\begin{figure}[ht]
  \includegraphics[scale =0.5]{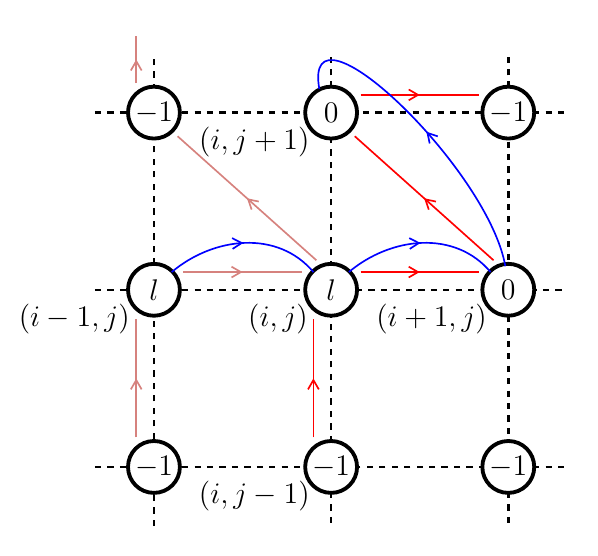}
  \caption{A schematic diagram to illustrate the case when lattice site 
	   being scanned is already labeled with a domain label.
	   As shown, the site $(i,j)$ that is being scanned is labeled 
	   with label $l$. In this case, the right and 
	   upper neighbors are checked for label $0$ and are added to the 
	   path of label $l$.}
  \label{old_dom}
\end{figure}


\subsubsection{Old domain}

Consider that the label of the lattice site is a positive integer $l$, 
indicating that $(i,j)$ is already identified as a member of the $l$th 
domain. In this case, no modification of the variables of the lattice site are
required. We then check the label of the right nearest-neighbor lattice 
site $(i+1,j)$. If the label is $-1$, then the scan continues to the upper 
nearest-neighbor lattice site $(i, j+1)$. On the other hand, if the label 
is $0$, then the lattice site belongs to the $l$th domain. So we have to 
update the domain and path variables to include $(i+1,j)$ as a
part of this domain. For this, we change the label of  $(i+1,j)$ to $l$ and 
link the end node of the $l$th path to the $(i+1,j)$ by setting 
$x^l_{\gamma^l,\delta^l}=i+1$ and $y^l_{\gamma^l,\delta^l}=j$. Then $(i+1,j)$ 
is made the end node by assigning $x^l_{i+1,j}=-2$ and $y^l_{i+1,j}=-2$ and 
then, update the end node variables to $\gamma^l = i+1$ and $\delta^l = j$.
The modifications associated with the addition of a node is complete by 
incrementing the node count ${\cal N}^l$ by one. The schematic diagram 
of this possibility is shown in the Fig.~\ref{old_dom}. For illustration, 
the path identified in the previous column scanning is shown in light red. 
In general, except for the label $l$ of site $(i,j)$, the three
cases (A, B and C) discussed earlier are applicable in the present case as
well. Case D is not applicable, as it corresponds to a domain of isolated
site. Like the earlier cases, the next step is to consider the possibility of
linking the upper neighbor lattice site $(i,j+1)$. It is to be noted that 
the four cases shown in Fig.~\ref{new_dom} and the case of the old 
domain discussed covers the possibilities encountered during the scanning 
of the sites.

Based on the discussions, in general, appending of a site $(i, j)$ as a new 
node to a path consists of three steps. First, the site $(i, j)$ is linked 
to the current end node of the path. Second, the site $(i, j)$ is identified 
as the end node. The addition of the node is completed by 
updating the end node variables and node count variable. These steps are 
common to all the cases.

\begin{figure}[ht]
  \includegraphics[scale=0.5]{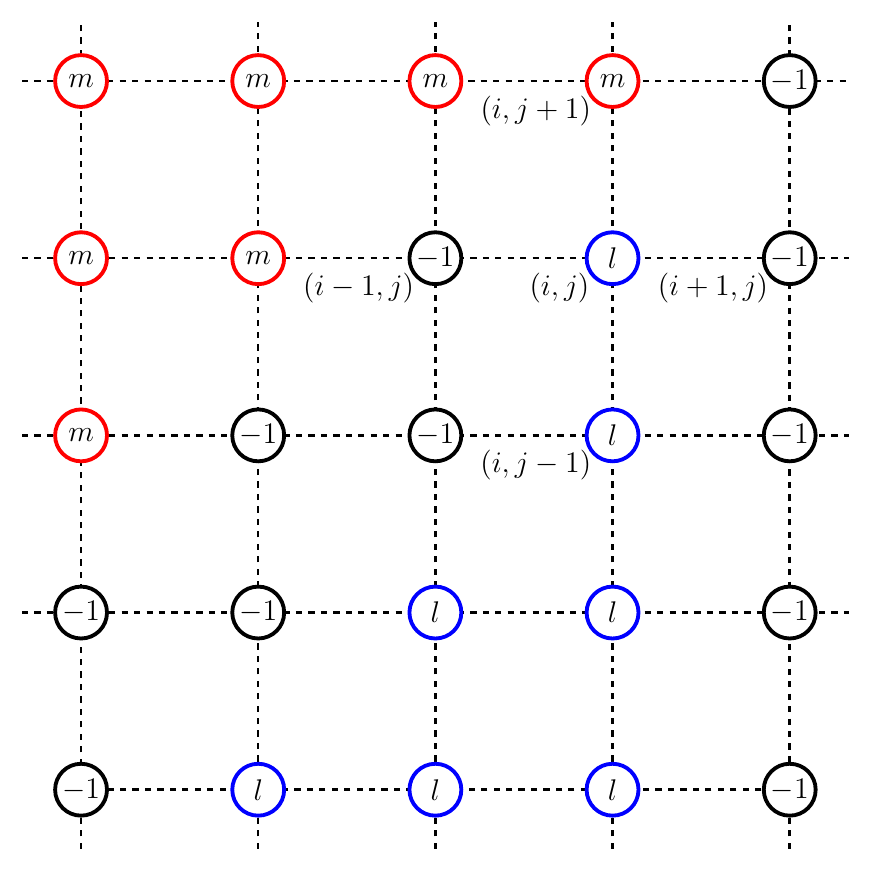}
  \caption{A schematic diagram to show the possibility of domain merging.
	   The neighboring sites $(i,j)$ and $(i,j+1)$ obtain different 
	   labels during the scan. In this case, the two domains need 
	   to be merged, and a single label is to be retained.}
  \label{merge_dom}
\end{figure}


\subsubsection{Domain merging}

 Merging of two domains occurs when the labels of the neighboring lattice 
sites along the column $(i,j)$ and $(i,j+1)$ are positive integers but 
different. As an example, consider the labels of these sites are $l$ and $m$, 
respectively, and assume $l<m$. A representative case of such a situation is
shown in Fig.~\ref{merge_dom}. Following the conventions adopted, the domain 
variables of the two are
$\{\alpha^l, \beta^l, x^l_{i,j}, y^l_{i,j}, \gamma^l, \delta^l, {\cal N}^l\}$ 
and
$\{\alpha^m, \beta^m, x^m_{i,j}, y^m_{i,j}, \gamma^m, \delta^m, {\cal N}^m\}$. 
To merge the two domains, the first step is to link the two corresponding paths 
and consolidate the two into a single one. For this the last node of the 
$l$th domain is linked to the first node of the $m$th node. 
It is done by setting $x^l_{\gamma^l,\delta^l}=\alpha^m$ and 
$y^l_{\gamma^l,\delta^l}=\beta^m$. As a convention, we tag the merged domain 
with the variables corresponding to the one with the lower label value, in 
this case $l$. The merger is complete by updating the last node variables as 
$\gamma^l=\gamma^m$ and $\delta^l=\delta^m$, and the total number of lattice 
sites in the domain ${\cal N}^l={\cal N}^l +{\cal N}^m$. As a last step, the 
$m$th domain is effectively nullified by setting ${\cal N}^m=0$.


\section{Charting the boundary}
\label{bndry}
 After we identify the domains, the next step is to map their boundaries. 
The boundary here means the outer edge of the domain. It defines the geometry
of the domain. It excludes the internal boundaries associated with 
voids within the domains. Determining the boundary is essential to 
investigate the properties of the domains and to apply the methods rooted in 
the percolation theory. The algorithms for determining the 
boundary are referred as the hull generating algorithms in the percolation 
theory \cite{ziff_84, ziff_89}.
To identify the boundary, we {\em march} along it in 
the clockwise direction, one bond at a time. For a domain, the starting 
site of the march $(x_s, y_s)$ is identified as the leftmost site along one 
of the rows. Then the march is initiated after identifying the hop along the 
boundary to reach $(x_s, y_s)$. We refer to this as  the {\em prior hop}.
\begin{figure}
  \includegraphics[scale = 0.5]{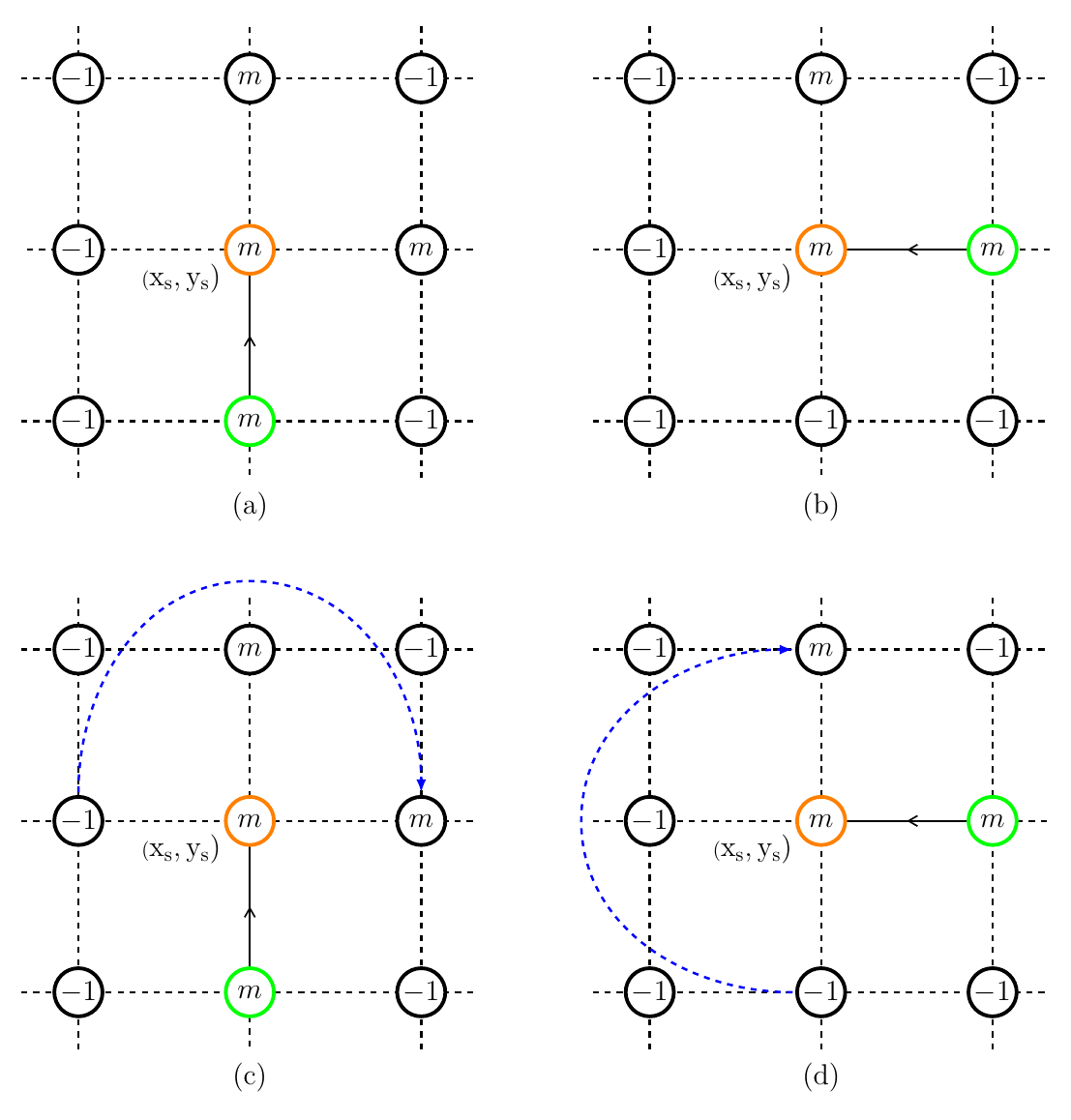}
  \caption{A schematic diagram to illustrate the prior hop and the 
	   clockwise scanning for the boundary {\em march}. 
	   The march 
	   is started from the leftmost site $(x_s,y_s)$ as highlighted by the 
	   peach color. The black arrow represents the prior hop to reach the 
	   site $(x_s,y_s)$. As illustrated in text, this prior hop can be 
	   either in upward direction [panel (a) or leftward direction
	   [panel (b)]. Panels (c) and (d), the blue dashed arrow indicates the 
	   scanning of the neighbors of $(x_s,y_s)$ in a clockwise manner
	   with respect to the orientation of the prior hop. The neighbour 
	   with label $m$ first encountered in this clockwise scan then becomes
	   next site on the boundary.}
  \label{bndry_left}
\end{figure}


\subsection{Identifying the prior hop}
\label{phop}
 The starting point $(x_s, y_s)$, as mentioned earlier, is the leftmost site 
of the domain along a row. Let the domain be labeled with $m$. Hence, there 
are only two possibilities of the 
prior hop through which it can reach the site. These are as shown in the 
Figs.~\ref{bndry_left}(a) and \ref{bndry_left}(b). Thus, it is sufficient to check the label
at the two lattice sites $(x_{s+1}, y_s)$ and $(x_s, y_{s-1})$. 
The one with label $m$ defines the originating site of the 
prior hop. With this information we can initiate the march. At the later 
hops too, the determination of the next hop requires the identification of 
the previous hop and this information is at hand once the march begins.


\subsection{Clockwise scan}
\label{clockscan}
 After the identification of the prior hop, we scan the other three
nearest neighbors of the site $(x_s, y_s)$ to identify the next site on the 
boundary. The scanning is done in the clockwise direction with respect to 
the orientation of the prior hop. This is schematically shown in 
Figs~.\ref{bndry_left}(c) and \ref{bndry_left}(d). The scan is terminated when 
we encounter a neighbor with label $m$. Let this neighbor be
identified as $(x_{s'}, y_{s'})$ and it is the next site on the boundary.
If all the three neighbors have label $-1$, then the march 
proceeds by retracing along the prior hop and origin of the prior hop is 
identified as the lattice site $(x_s',y_s')$. The bond connecting the two 
lattice sites $(x_s, y_s)$ and $(x_{s'}, y_{s'})$ defines the orientation of 
the prior hop to scan for the next lattice site on the boundary after 
$(x_{s'}, y_{s'})$. This process of scanning is repeated until we return to 
the starting site $(x_s, y_s)$.

\begin{figure}[ht]
  \centering
  \includegraphics[scale = 0.5]{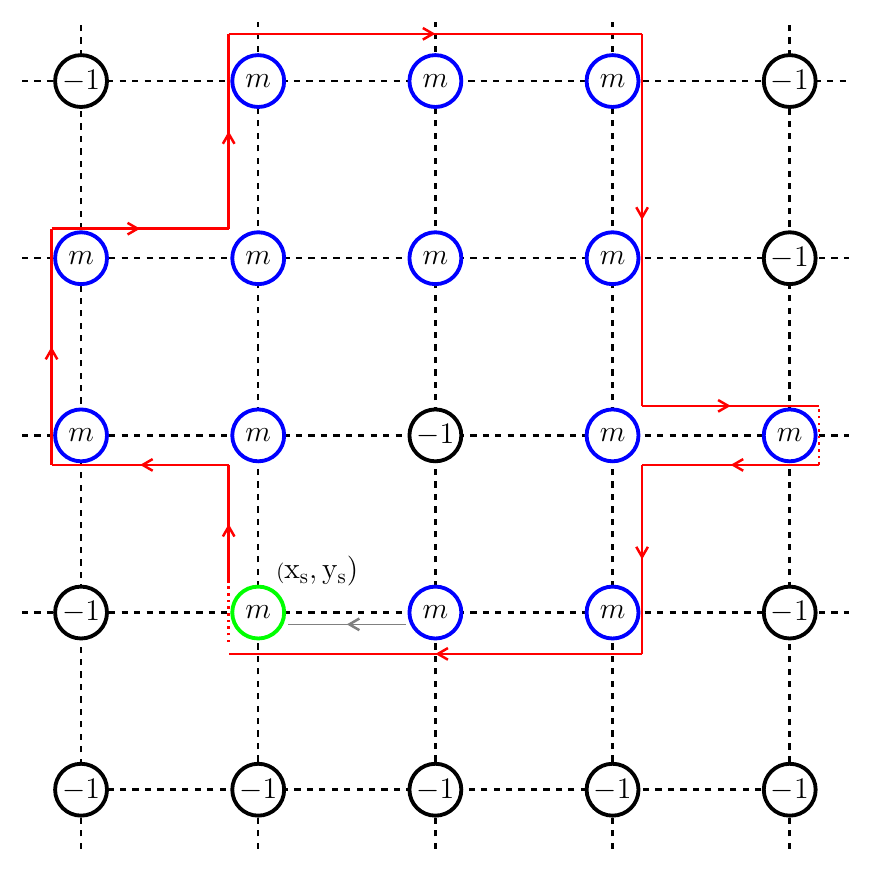}
    \caption{A schematic illustration of the march along the boundary of a 
	     cluster of label $m$. The leftmost site $(x_s,y_s)$ in the bottom
	     row is the starting site of the march.   
	     The prior hop along the boundary to reach  
	     $(x_s,y_s)$ is the site $(x_{s+1},y_s)$, thereby this is leftward
	     prior hop. This is shown by the gray colored arrow. The red arrows 
	     denote the directionality of the boundary march of the cluster
	     shown in the figure.}
  \label{bndry_march}
\end{figure}

As a representative case, we show the path along the boundary of a cluster in 
Fig.\ref{bndry_march}. The starting site $(x_s,y_s)$ is the leftmost site of
the bottom row and is highlighted in green color. The prior hop to 
reach $(x_s,y_s)$ is leftward in this case and shown by a gray 
arrow. Once the starting site and the prior hop are identified, the 
scanning of nearest neighbors proceeds to identify the next site on the
boundary. As explained, the march is completed when we return to the 
initial starting site $(x_s,y_s)$. An important prerequisite for the boundary 
march is thus the identification of the starting site and the prior hop.


\section{Comparison of the algorithm with other methods}
\label{comparison_section}
As mentioned in the Introduction, there are other well-known algorithms to 
identify domains in the percolation analysis. In this section, we compare the 
proposed algorithm with two of the standard algorithms; these are the HK and 
the recursive neighbor search algorithms. The comparisons are based on two
parameters, the compute time and memory required. To get the 
general trends, the system size is varied over four orders of magnitude from 
$10^4$ to $10^8$. And, for each system size the runtime is taken as the 
average of $40$ configurations. The configurations are generated using 
univariate random numbers generated using the Marsenne twister pseudo 
random number generator.


\subsection{Hoshen-Kopelman algorithm}

\begin{figure}[ht]
  \centering
  \includegraphics[scale = 0.5]{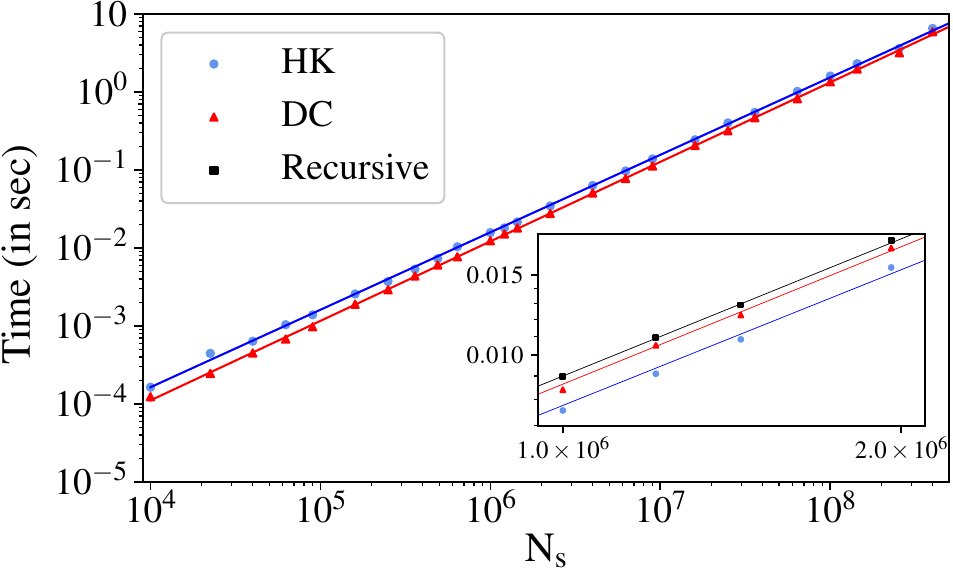}
  \caption{The runtime of the HK (solid blue circles), the proposed DC 
           algorithm (solid red triangles), and recursive neighbor 
           search algorithms (solid black squares) for different lattice 
           sizes $N_s$. The main panel shows total time required for the first 
           scan and the postprocessing in the HK and the DC algorithms. For
           comparison of all the three methods, the inset plot shows the time 
           required for the first scan in the three algorithms.}
  \label{time_comp}
\end{figure}

It is a multiple cluster labeling algorithm widely used in clustering and 
percolation studies. In this algorithm, the lattice sites and their 
neighbors are scanned systematically, and the sites are labeled according to 
the occupancy. So in this method a cluster may comprise smaller clusters 
or subclusters with different labels. An equivalence class of the 
different labels of the subclusters identify a cluster. Thus, the HK algorithm 
constructs a coarse-grained linking between the cluster labels. On the 
contrary, the  DC algorithm is fine-grained in character as it creates 
intersite links. Such a fine-grained approach may be better 
suited for the characterization and properties calculations using percolation 
theory. In the HK framework, additional lattice scan is required while 
computing cluster properties. However, in the DC algorithm, the linked-list 
identifies members of a cluster and traverse through the cluster with a unique
path. Additional scans are not required. Thus, the DC algorithm 
is suitable for percolation problems that rely on identifying and analyzing 
the cluster properties.

Furthermore, an additional scan after the first lattice scan is required in 
the HK algorithm to label the clusters with a unique cluster label. This 
postprocessing step in the HK algorithm makes it slower than the DC 
algorithm, as shown in Fig.~\ref{time_comp} (main panel). The details of the 
comparative studies related to the Fig.~\ref{time_comp} are given in  
Appendix \ref{appendA}. 

The memory required in the DC algorithm is about three times larger than the 
HK algorithm. This is expected as the DC method needs memory to construct the 
linked lists. Considering the recent advances in memory hardware technology 
this is not a  significant limitation. However, a decade ago, when very large 
memory configurations were not possible, it would have made a huge difference. 
In summary, the fine-grained linking of the DC algorithm offers 
an advantage and renders it suitable for the cluster characterization and 
properties calculations.


\subsection{Recursive neighbor search algorithm}
The recursive neighbor search algorithm involves scanning and labeling the 
neighbors of the occupied sites recursively. In this method, if an occupied 
site is found during the scan, then it is relabeled and the neighbors of this 
site are scanned. This process is continued recursively until all the sites of 
the cluster are identified. The method is implemented with relative ease using 
recursive function call. The key advantage of this approach is only the 
required cluster can be identified without the need to identify other clusters. 
However, as shown in the inset of Fig.~\ref{time_comp} and discussed in 
Appendix \ref{appendA}, this algorithm is slower than the DC and the HK 
algorithms. In addition, studies have shown that the number of the
recursive functions becomes very large for larger system 
sizes \cite{herrero_07}. This limitation does not apply to the DC 
algorithm as it is not based on recursion. Further, the recursive neighbor 
search is also not efficient for properties calculations as the domain 
information is not stored and each cluster characterization requires a 
lattice scan. It is to be noted that the problems studied in the 
percolation theory are not restricted to cluster identification and their 
characterizations. The choice of the numerical approach to study the 
percolation problems should therefore be based on the nature and 
complexity of the problem of interest.


\section{MI-SF quench dynamics}
\label{application}
The MI phase to the SF phase quench dynamics of the Bose-Hubbard
model \cite{fisher_89}, a model which describes the physics of ultracold 
bosons in optical lattices \cite{jaksch_98}, has been experimentally 
realized \cite{braun_15}. During the quench, as the system crosses the 
critical point, SF domains are formed within MI as a background.
We quantify the number of the SF domains using our domain counting algorithm. 
To discuss the dynamics, we first introduce the BHM Hamiltonian.


\subsection{BHM Hamiltonian}
 The BHM Hamiltonian which describes the physics of ultracold bosonic atoms 
loaded in a 2D square optical lattice is \cite{fisher_89, jaksch_98}
\begin{equation}
  \hat{H} = -\sum_{\langle i,j \rangle}J\left( \hat{b}_{i}^{\dagger}
               \hat{b}_{j} + {\rm H.c.} \right)
              + \sum_{i}\hat{n}_{i} \left [\frac{U}{2}(\hat{n}_{i}
              -1)- \mu \right],
  \label{bhm}
\end{equation}
where $i \equiv (p,q)$ represent the lattice indexes, and $j \equiv (p',q')$ 
are the indexes of it's neighboring lattice site,
$\hat{b}_{i}^{\dagger}$ ($\hat{b}_{i}$) are the creation (annihilation)
operators, $\hat{n}_{i}$ is the bosonic occupation number operator, and the
summation indexes within $\langle\cdots\rangle$ denote the sum over the
nearest neighbors. Further, $J$ is the hopping strength, $U>0$ is the
on-site inter-atomic interaction strength, and $\mu$ is the chemical potential.

The incompressible MI state and the compressible SF state are the two ground 
states of the BHM Hamiltonian in the strongly interacting ($J/U \ll 1$) and 
weakly interacting ($J/U \gg 1$) domains, 
respectively \cite{fisher_89, jaksch_98}. The quantum phase transition 
between these two phases has been observed experimentally
\cite{greiner_02_1, greiner_02_2}. The MI state has integer commensurate 
lattice site occupancies, and the bosons are pinned to the lattice sites. 
The SF state, on the other hand, features a real valued occupancy, and  it is
a conducting phase. The two phases are identified based on the SF order
parameter 
\begin{equation}
 \phi_{p,q} = \langle \hat{b}_{p,q} \rangle.
\end{equation}
It is zero in the MI phase and nonzero in the SF phase. For a homogeneous 
lattice system, $\phi_{p,q}$ is uniform throughout the lattice. 
For our studies, we use the single-site Gutzwiller mean-field (SGMF) method to
obtain the ground state of the model and phase diagram. In this method, the 
annihilation (creation) operators in Eq.~(\ref{bhm}) are separated into a 
mean-field $\phi$ ($\phi^{*}$) and a fluctuation 
operator \cite{rokhsar_91, sheshadri_93}.  The Hamiltonian in Eq.(\ref{bhm}) 
can then be approximated as the mean-field Hamiltonian which is a sum of 
single site Hamiltonians $\hat{H}_{\text{MF}}=\sum_{p,q}\hat{h}_{p,q}$. We 
perform self-consistent calculation of $\phi_{p,q}$ until the desired 
convergence is obtained. The details of using this method in our computations 
are given in our previous 
works \cite{bai_18, pal_19, bandyopadhyay_19, suthar_20_1, bai_20, suthar_20_2, 
sable_21}.


\subsection{Quench dynamics}

To study the MI-SF quench dynamics of the system,  the hopping amplitude
$J$ is ramped from an initial value $J_i$ to a final value
$J_f$. These are chosen such that $J_i$ and $J_f$ correspond to the MI and SF
phases, respectively. The remaining parameters of the system are held
fixed. Then, the temporal evolution of the system during the quench and 
afterwards is described by the time-dependent Schr\"{o}dinger equation
\begin{equation}
   i\hbar\partial_t \Ket{\psi}_{p,q} = \hat{h}_{p,q} \Ket{\psi}_{p,q},
   \label{schdinger_eqn}
\end{equation}
where $\Ket{\psi}_{p,q}$ is the wavefunction at site $(p,q)$. Due to the 
intersite coupling through the order parameter $\phi$, we obtain 
a set of coupled partial differential equations. These are solved using
the fourth-order Runge-Kutta method. To start the quench, we obtain the
equilibrium wavefunction with the $J = J_i$ and introduce  phase
and density fluctuations to it \cite{sable_21}. These fluctuations simulate 
the quantum fluctuations essential to drive the quantum phase transition.
To calculate system properties we take the ensemble average of a set consisting 
of 80 such randomized initial states.

We examine the nonequilibrium dynamics of the system during the quench 
from the KZM perspective. It categorizes the quench dynamics
into three temporal regimes \cite{damski_05, damski_06}, corresponding to
the adiabatic, impulse, and adiabatic regime. These temporal regimes arise 
due to the critical slowing down near the quantum critical point (QCP).
It predicts the rate of the topological defects formation during the course
of the quench dynamics \cite{kibble_76, kibble_80,zurek_85, zurek_96, 
delCampo_14}. These defects are generated at the meeting points of the
domains of the symmetry broken phase. This is is due to the local 
gauge choices of the order parameter associated with the domains. For point 
defects like vortices, the density of defects is hence proportional to the 
number of domains. It is to be mentioned that the transition from MI to SF 
phase breaks the global $U(1)$ symmetry spontaneously. Then number of the 
domains $N_D$ satisfies the scaling law
\begin{equation}
        N_D \propto \tau_{Q}^{-d},
\end{equation}
where $1/\tau_{Q}$ is the quench rate and exponent $d =2\nu/(1 + \nu z)$. 
Here $\nu$ is 
the critical exponent of the equilibrium correlation length
and $z$ is the dynamical critical exponent. It is to be noted that these 
scaling laws are applicable at $\hat{t}$, which is the time at which
the system transits from the impulse to the adiabatic domain. The details 
of locating $\hat{t}$ are given in our previous work \cite{sable_21}. The
same scaling law is also applicable to the defect density $N_v$. For the 
MI-SF transition the defects are the vortices and their density is 
given by \cite{shimizu_misf_18, shimizu_dwss_18, shimizu_dwsf_18, sable_21}
\begin{equation}
  N_v = \sum_{p,q} |\Omega_{p,q}|, 
  \label{vort_den}
\end{equation}
with
\begin{eqnarray}
  \!\!\!\!\!\!\!\!
  \Omega_{p,q} &=& \frac{1}{4}\big [\sin(\theta_{p+1,q} - \theta_{p,q})
                   + \sin(\theta_{p+1,q+1} - \theta_{p+1,q})  
                                         \nonumber \\
               &&  -\sin(\theta_{p+1,q+1} - \theta_{p,q+1}) - 
                   \sin(\theta_{p,q+1} - \theta_{p,q})\big],
  \label{vort_def}
\end{eqnarray}
where $\theta_{p,q}$ is the phase of $\phi_{p,q}$. In the results section,
we first calculate the critical exponent $d$ from the defect density $N_v$.
Then we use our method to calculate $N_D$ and show that we get similar value
of the critical exponent based on $N_D$. This serves as an excellent cross 
checking of two different approaches to estimate the same critical exponent.


\subsection{Results}
To study the MI-SF nonequilibrium quench dynamics, we consider a system of 
size of $100 \times 100$. The Hamiltonian is scaled with $U$ and time is 
defined in the units of $\hbar/U$. The hopping amplitude $J$ is evolved
using the following quench protocol:
\begin{equation}
  J(t) = J_i + \frac{(J_c - J_i)}{\tau_Q}(t + \tau_Q).
\end{equation}
With this protocol, we have, $J(-\tau_Q) = J_i$ and  $J(0) = J_c$ is the
QCP of the MI-SF phase transition. For our study we take $J_i = 0.0U$ and 
$J_f = 0.08U$ and fix the chemical potential $\mu = 0.41U$. The value of 
the $\mu$ is chosen so that it corresponds to  the tip of the MI(1) lobe 
and $J_c = 0.042U$.  Thus, at $t = -\tau_Q$, the system is in the MI(1) 
phase, and at $t = t_f$, it is in the SF phase when the quantum quench ends.
\begin{figure}[ht]
  \includegraphics[height = 3.5cm]{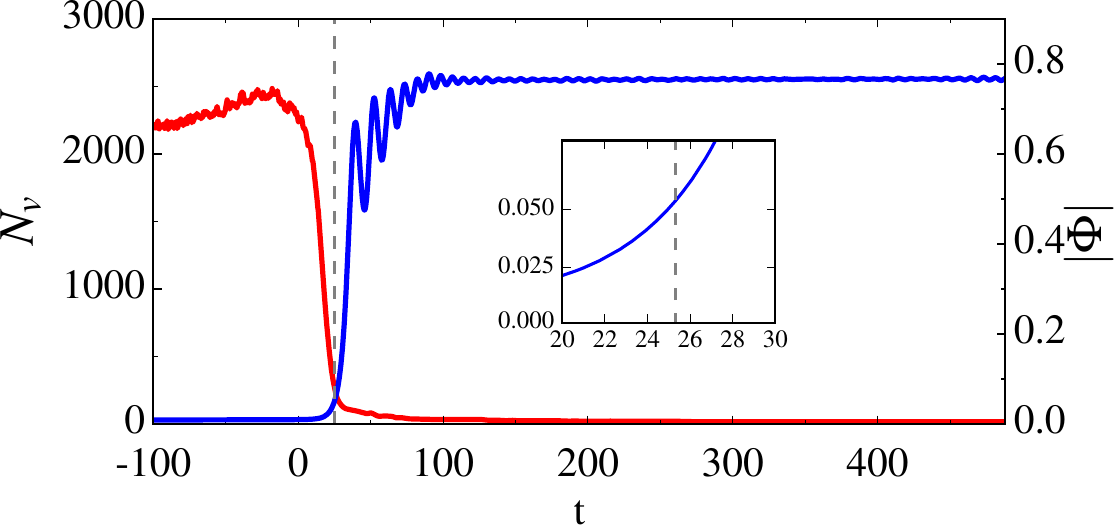}
  \caption{The time evolution of the vortex density $N_v$ (red) and 
	   $|\Phi|$ (blue) as a function of time, for $\tau_Q = 100$. 
	   The QCP is 
	   crossed at $t = 0$, and the dashed gray line indicates the time 
	   $\hat{t}$. The order parameter $|\Phi|$ rises exponentially after
	   $\hat{t}$, as shown in the inset. The vortex density $N_v$ exhibits
	   a steep decrease after crossing the QCP due to the annihilation
	   of the vortex-antivortex pairs.}
  \label{phi_nv}
\end{figure}


\subsubsection{Transition from MI to SF phase}
As an indicator of the quench dynamics, the temporal evolution of
$|\Phi|=\sum_{p,q}|\phi_{p,q}|/N_s$ and $N_v$ during the quench, for 
$\tau_Q = 100$ are shown in Fig. \ref{phi_nv}. Here $N_s$ denotes the number 
of lattice sites. In the initial stages of the quench dynamics, $|\Phi|$ is 
close to zero ($\sim 10^{-3}$) and remains so until $\hat{t}$. In the 
equilibrium MI state, $|\Phi|$ is zero, but in the quench dynamics it is finite 
due to the fluctuations added in the initial-state preparation. As the QCP is 
crossed at $t=0$, the system ought to evolve into SF phase ($t>0$) and acquire 
a larger $|\Phi|$. But, as the system is still in the impulse domain, the 
$|\Phi|$ remains small until it ends at $\hat{t}$. Post $\hat{t}$, there is 
an exponential increase of $|\Phi|$, which is discernible from the plot in 
Fig. \ref{phi_nv}. After the exponential increase, the $|\Phi|$ settles to 
a steady-state value.
\begin{figure}[ht]
  \includegraphics[width = 8.5cm]{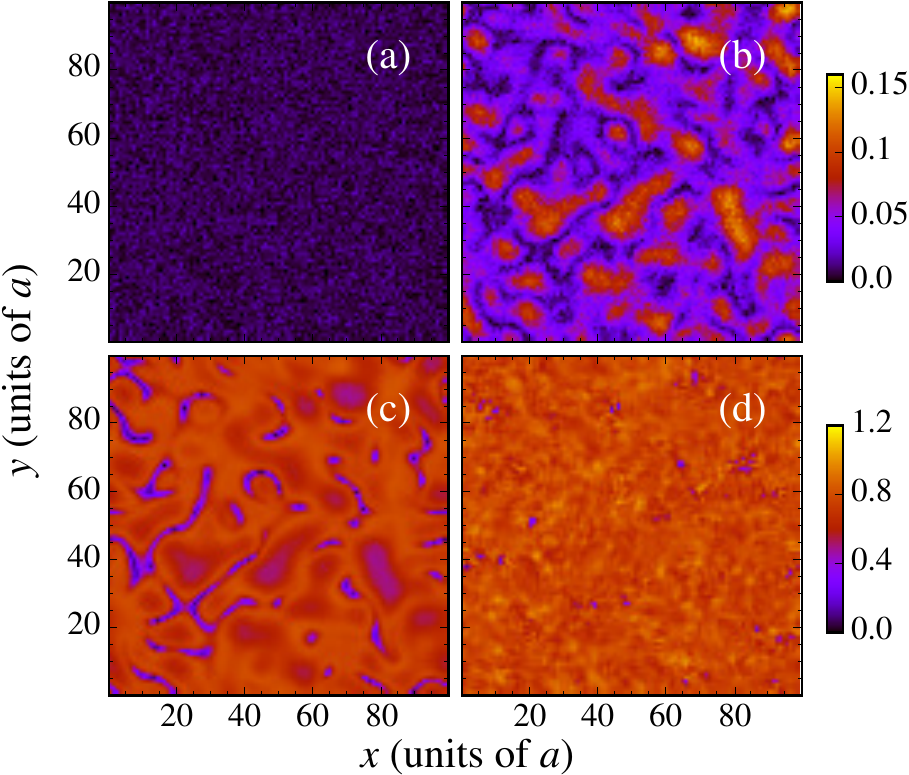}
  \caption{Snapshots of $|\phi_{p,q}|$ at certain time instants for
           $\tau_Q = 100$. At initial time $t = -\tau_Q$, $|\phi_{p,q}|$ is 
	   small as shown in (a). After $\hat{t}$, the $|\phi_{p,q}|$ 
	   increases and domains of SF are formed. This is shown in panel (b)
	   which is at $t = \hat{t}$. These domains disappear due to the 
	   merging as time progresses and the system becomes homogeneous, 
	   as shown in panels (c) and (d).}
  \label{phi_snap}
\end{figure}

The snapshots of $|\phi_{p,q}|$ at different times are shown in 
Fig. \ref{phi_snap}. The small values of $|\phi_{p,q}|$ at initial time 
$t = -\tau_Q$ is as shown in Fig. \ref{phi_snap}(a). The figure also indicates 
the fluctuations present in the values of $|\phi_{p,q}|$. The 
Fig. \ref{phi_snap}(b) shows the formation SF domains at $\hat{t}$ when the
system re-enter the adiabatic domain. The domains can be easily counted
using our method. When $J/U$ is further increased, these domains grow in size, 
and merge through the phase ordering process. This is visible from the
Figs. \ref{phi_snap} (c) and \ref{phi_snap}(d), in these figures 
$|\phi_{p,q}|$ is almost uniform.

The evolution of defect density $N_v$ is complimentary to that of 
$|\Phi|$. At the initial stages of the quench, $N_v$ is high 
($\approx 2200$), this is due to the large phase fluctuations
added to the initial state. It decrease after the system crosses QCP. This
happens as the SF domains begin to form and phase coherence within the domains
prevent the presence of a vortex inside a domain. As time progresses, the 
phase ordering takes places, and these domains merge. The domain merger 
results in the annihilation of the vortex-antivortex pairs, further reducing 
$N_v$. This is discernible from Fig. \ref{phi_nv}.


\subsubsection{Critical exponents and scaling laws}

\begin{figure}[ht]
  \centering
  \includegraphics[height = 6.3cm]{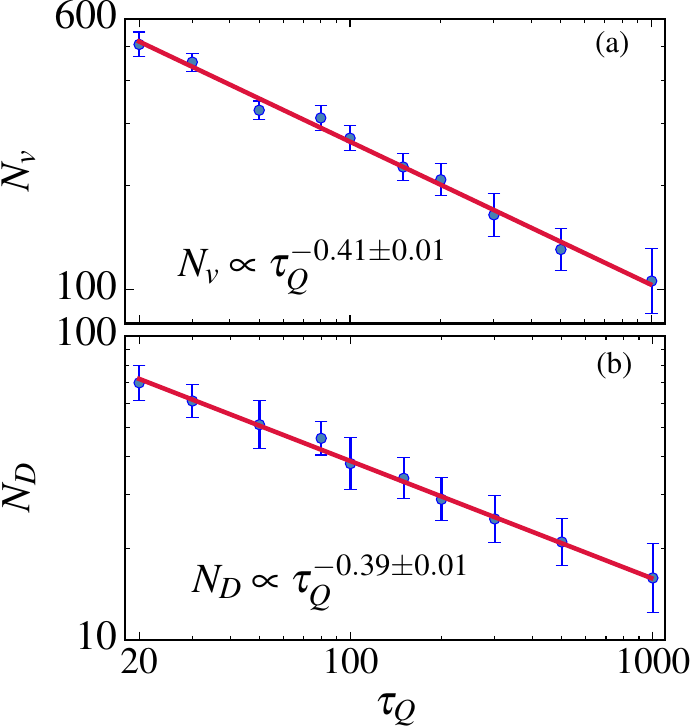}
    \caption{Scaling of the vortex density $N_v$ and number of domains
           $N_D$, with respect to $\tau_Q$. From (a), we see that
           the exponent $d = 0.41$. The scaling exponent for $N_D$ versus 
	   $\tau_Q$ is $0.39$, as shown in (b).
	   The blue error bars represent the standard deviation of the
           data values.}
  \label{nv_scal}
\end{figure}

To study the scaling of the $N_v$ with $\tau_Q$, we compute $N_v$ over a 
range of $\tau_Q$. As mentioned earlier, the scaling laws are considered at
time $\hat{t}$. Hence, we compute $N_v$ at $\hat{t}$ for every $\tau_Q$. The  
log-log plot of the values obtained are shown in Fig. \ref{nv_scal}(a). From
least-squares fitting, we get the value of the critical exponent
$d$ as $0.41$. That is, $N_v \propto \tau_Q^{-0.41}$.

Following similar analysis, we determine the scaling exponent of the 
$N_D$ with the quench rate. As shown in Fig.\ref{phi_snap} (b), SF domains 
begins to appear at $\hat{t}$. These domains are characterized by
a finite value of the SF order parameter. Since the MI regions also have a
small nonzero SF order parameter, owing to the initial fluctuations, we set a
threshold $\epsilon$ of the $|\phi_{p,q}|$ to distinguish the SF phase from the
MI phase. The value of $\epsilon$ is taken as the average of $|\phi_{p,q}|$ 
over the prominent MI phase regions in the system at $t=0$. In our 
computations, based on this definition,  we get $\epsilon \approx 0.07$. We 
then calculate the ensemble-averaged value of $N_D$ using our method. From 
the results, the scaling of $N_D$ with $\tau_Q$ is as shown in 
Fig. \ref{nv_scal}(b). We observe a power-law scaling, 
\begin{equation}
  N_D\propto \tau_Q^{-0.39},
\end{equation}
that is, the critical exponent $d=0.39$. This implies that the scaling of 
the $N_D$ with $\tau_Q$ is approximately same as the scaling of
the $N_v$ with $\tau_Q$. This is expected from the KZM, as the density of the
topological vortices is used as a proxy for the number of domains. Thus,
the domain counting algorithm predicts an exponent that is in a good agreement
with the exponent obtained from the vortex density. Here it is to be noted
that the methods used in computing $N_D$ and $N_v$ are different. One is
based on the current distribution and our method is based on the 
identification of clusters. The two in essence serve as independent checks
of the KZM scaling of the MI-SF transition.


\section{Disordered Bose-Hubbard model}
\label{percol_bg}

  The analysis of the MI-SF quench dynamics utilized the domain counting
aspect of the method we have developed. The fine-grained nature of the 
method also makes it suitable for detailed analysis of the clusters as well.
We apply this to study the critical properties of the BG to SF transition 
in the 2D square optical lattice with disorder. The system is modeled with 
the disordered Bose-Hubbard model (DBHM). Earlier studies have identified
the ground-state phase diagram of the DBHM and have shown the absence of a 
direct MI-to-SF transition \cite{fisher_89, buonsante_07, 
pollet_09, bissbort_10, pisarski_11, soyler_11, pal_19}.
The MI-SF transition is intervened by the BG phase, which is 
characterized by nonzero compressibility and zero superfluid stiffness. 
Structurally, this is essentially the MI phase inlaid with SF islands and the
SF islands lead to the finite compressibility. However, the superfluid
stiffness is zero as the islands cannot generate phase coherence across 
the system. Thus, in terms of percolation theory, the SF islands in the 
BG phase are nonpercolating. But, the clusters percolate when the BG phase 
undergoes a transition to the SF phase. Therefore, the BG-to-SF transition 
can be viewed as the percolation analysis of the SF clusters. Previous 
works \cite{niederle_13, nabi_16, barman_13} have studied the geometric 
properties of the SF clusters and compared the onset of superfluidity with 
the percolating transition of the system. In this work, we delve into the 
cluster properties using our method. In particular, we study the power-law 
divergence of the correlation length of the system near the percolating 
transition and extract the critical exponent $\nu$ quantifying the 
divergence. We also compute the fractal dimension of the hulls 
of the SF clusters using the boundary walk algorithm discussed earlier.
These studies reveal that the BG-to-SF transition falls in the 
universality class of 2D random percolation.


\subsection{DBHM Hamiltonian} \label{dbhm_model}
 The DBHM Hamiltonian of the 2D square optical lattice is 
\cite{fisher_89, niederle_13, pal_19}
\begin{equation}
  \hat{H} = -\sum_{\langle i,j \rangle}J\left( \hat{b}_{i}^{\dagger}
               \hat{b}_{j} + {\rm H.c.} \right)
              + \sum_{i}\hat{n}_{i} \left [\frac{U}{2}(\hat{n}_{i}
              -1)- \tilde{\mu}_{i} \right],
  \label{dbhm}
\end{equation}
where, except for $\tilde{\mu}_i$, the model parameters have the same meaning 
as the BHM Hamiltonian in Eq.(\ref{bhm}). Here the effective chemical 
potential $\tilde{\mu}_i=\mu - \epsilon_i$ is site-dependent, and  
$\epsilon_i\in[-\Delta, \Delta]$ is a univariate random number to simulate 
diagonal disorder. The parameter $\Delta$ denotes the strength of the 
disorder. In the experiments, the disordered lattice potential is generated by 
shining a speckle beam \cite{clement_05,clement_08, white_09}.

 The ground-state phase diagram of the DBHM exhibits the quantum phases
determined by the competition between the hopping energy $J$, the interaction
energy $U$, and the disorder strength $\Delta$. Then, like in the BHM, we use 
the SGMF method to obtain the quantum phases and their characteristic 
order parameters. For a moderate $\Delta$, when $J/U$ is small, the strong 
onsite repulsion favors the incompressible MI phase and atoms are pinned 
to the sites. For large $J/U$, the large hopping strength favors the 
compressible SF phase and atoms are itinerant. In the intermediate $J/U$ the 
two phases are separated by the BG phase. As mentioned earlier, the BG phase is 
characterized by SF islands in the background sea of insulating MI phase. 
These islands impart finite number fluctuations $\delta n_{p,q}$ to the BG 
phase. Hence, $\delta n_{p,q}$ is an order parameter to distinguish the BG 
phase from the number coherent MI phase. For the site $(p,q)$,
\begin{equation}
  \delta n_{p,q} = \sqrt{\langle\hat{n}_{p,q}^2\rangle 
                   - \langle \hat{n}_{p,q} \rangle^2},
  \label{nfluc}
\end{equation}
where the expectation value $\langle .. \rangle$ is taken with respect to the
ground state. At the BG-SF phase boundary, the SF stiffness $\rho_s$ is the
relevant order parameter to differentiate the BG and SF phases.
It denotes the finite energy required to alter the phase of the wavefunction 
of the system. Since the SF phase is phase coherent, it exhibits a stiffness 
or resistance for the phase change. So, $\rho_s$ is nonzero in the SF phase. 
But, it is zero in the MI phase and nonzero but small in the BG phase.
To compute $\rho_s$, we impose twisted boundary condition.
It modifies the hopping terms by introducing Peierls phase factors 
\cite{byers_61, shastry_90, roth_03, gerster_16}. For the DBHM, we impose 
twisted boundary condition along the $x$ direction and transform the
hopping terms as \cite{roth_03, gerster_16}
\begin{equation}
  J\left( \hat{b}_{p+1,q}^{\dagger} \hat{b}_{p,q} + {\rm H.c.} \right) 
  \rightarrow J\left( \hat{b}_{p+1,q}^{\dagger} \hat{b}_{p,q} 
  e^{i2\pi\varphi/L_x} + {\rm H.c.} \right).
\end{equation}
The SF stiffness is then defined as
\begin{equation}
  \rho_s=\frac{L_x}{8\pi^2}\frac{\partial^2E_0}{\partial\varphi^2}
	 \Big|_{\varphi = 0},
  \label{sfstiff}
\end{equation}
where $E_0$ is the ground-state energy with twisted boundary condition.

 For our studies based on the statistics of the SF clusters in the BG
phase, we choose the system size as $1000 \times 1000$. The order parameters
$\delta n_{p,q}$ and  $\rho_s$ are taken as ensemble average of 60 disorder 
realizations. Here it is to be added that we observe finite-size effects 
for system sizes up to $100 \times 100$.
We attribute this to the poor statistics as a result of fewer domains in a 
smaller system size. The remedy is to increase the number of domains by
increasing the system size and improve the statistics. This reduces the 
sample to sample variation in the quantities like the correlation length.
The system size of $100 \times 100$ is, however, suitable for other studies 
to probe properties which depend on the average or coarse grained measures 
like the determination of the phase diagram and quench dynamics.


\subsection{Percolation analysis}
The percolation theory analyzes the statistical and geometrical properties 
of the clusters of sites on a lattice \cite{stauffer_92}. Specifically, in 
the site-percolation problem, every lattice site is independently and 
randomly occupied with probability $p$. The collection of neighboring 
occupied sites is termed as a cluster. For small $p$, a majority of the 
clusters constitute small number of sites and are isolated, while for large 
$p$, a majority of the occupied sites form a percolating cluster which spans 
from one edge to the opposite edge. Thus, there exists a critical threshold 
probability $p_c$ so that for $p < p_c$, there is no spanning cluster, and 
for $p \ge p_c$, there exists at least one spanning cluster. Such a 
percolating transition is characterized by power-law divergences
and universal critical exponents. One property which exhibits divergence 
at transition is the correlation length of the system $\xi$. It can be 
defined as an average over the cluster radii in the system
\begin{equation}
  \xi^2 = \frac{ \sum_s R^2_s s^2 n_s}{\sum_s s^2 n_s},
  \label{corrlen}
\end{equation}
where $R_s$ is the gyration radius of the cluster of $s$ sites and the
$n_s$ denotes the average number of clusters of size $s$ per site. In the
summation, the contribution from the infinite, percolating clusters is omitted
\cite{stauffer_92, sahimi_21}. At the percolation transition $\xi$
shows power-law divergence,
\begin{eqnarray}
  \xi \propto |p - p_c|^{-\nu}.
  \label{corrlen_scal}
\end{eqnarray}
The calculation of $R_s$ involves the evaluation of the distance of a 
lattice site from the center of mass of the 
cluster \cite{stauffer_92, sahimi_21}. For systems with the periodic boundary 
conditions (PBC), due to the absence of edges, there are two possible 
definitions of the distance between two lattice sites. To resolve this 
in a consistent way, we unwrap the system. So that the clusters in the
original system are mapped onto a system where dimensions are doubled. In 
the larger system, the distance between the lattice sites in a cluster are
defined without ambiguity. The details of this procedure are given in 
the Appendix \ref{appendB}.

In the BG phase, the sites with nonzero SF order parameter are
considered as occupied sites. Hence, as the critical $J_c$ of the
BG-to-SF transition is approached, we expect the SF clusters to percolate.
So, we can equivalently write the power law divergence in terms of 
reduced hopping strength as
\begin{equation}
  \xi \propto |J - J_c|^{-\nu}.
  \label{corrlen_scalJ}
\end{equation}
To identify the occupied sites we choose a threshold of $10^{-3}$ for the
SF order parameter.


\subsection{Results of BG-to-SF transition}
\begin{figure}[ht]
  \includegraphics[width = 8.5cm]{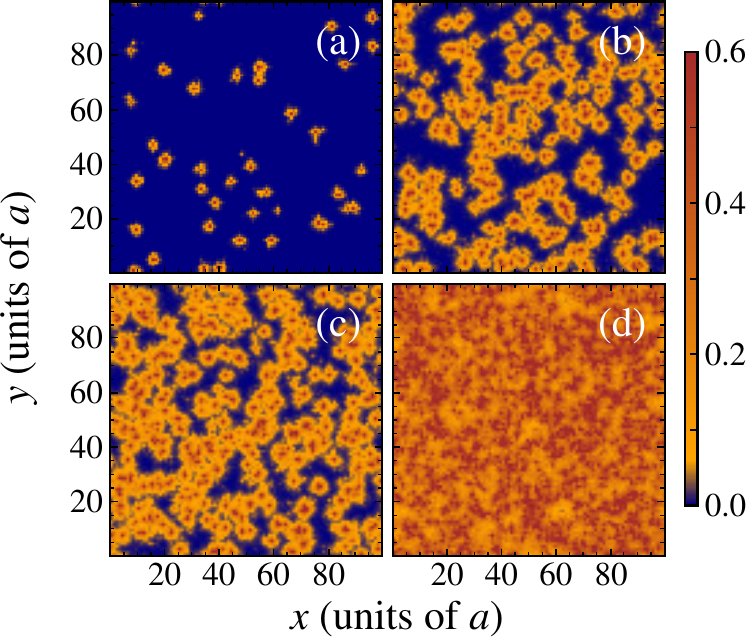}
  \caption{Snapshots of $|\phi_{p,q}|$ at different $J/U$ values. Panels
	(a)-(d) correspond to the $J/U$ values as $0.009$, $0.017$, $0.019$,
	and $0.025$ respectively. As the $J/U$ is increased, the SF domains
	percolate and the system undergoes a transition from the BG to SF phase.
	The chosen parameters are $\mu = 0.2U$ and 
	$\Delta = 1.2U$. Panel (b) illustrates the order parameter 
	profile just before the percolation transition.}	
  \label{percol_phi}
\end{figure}

For the percolation analysis we take $\Delta = 1.2U$ and $\mu = 0.2U$ and 
scan the phases as a function of $J/U$. The phase diagram in the 
$J/U -\mu/U$ plane, as reported in Ref.~\cite{pal_19}, constitutes the BG and 
SF phases. The absence of the MI phase is due to the high disorder strength.
The snapshots of $|\phi_{p,q}|$ for selected values of $J/U$ are shown in 
Fig. \ref{percol_phi}, for a $100\times100$ system and $\mu/U=0.2$. In the 
figure the rare SF islands in the BG phase are visible in the 
Fig.~\ref{percol_phi}(a) for $J/U = 0.009$. These SF islands are 
nonpercolating, and hence the introduction of a phase twist does not cost 
energy. The SF islands increase in size as $J/U$ is increased; this is 
evident in Fig.~\ref{percol_phi}(b), which shows $|\phi_{p,q}|$ for  
$J/U = 0.017$. This snapshot illustrates the order parameter 
profile before the percolation transition and highlights the large SF islands 
in the system. Although the SF islands are large, there is no spanning cluster.
On reaching critical $J/U$ a spanning cluster emerges and system undergoes
percolation transition. This is illustrated in the Fig.~\ref{percol_phi}(c),
the system supports a spanning cluster for $J/U = 0.019$. Due to the spanning
SF cluster, introducing a phase twist costs energy and superfluid stiffness
$\rho_s$ assumes a finite value. On further increase of $J/U$ the background 
MI phase region is completely depleted and the entire system is in the SF 
phase.  This is discernible from the Fig.~\ref{percol_phi}(d).

\begin{figure}[ht]
  \centering
  \includegraphics[height = 7.3cm]{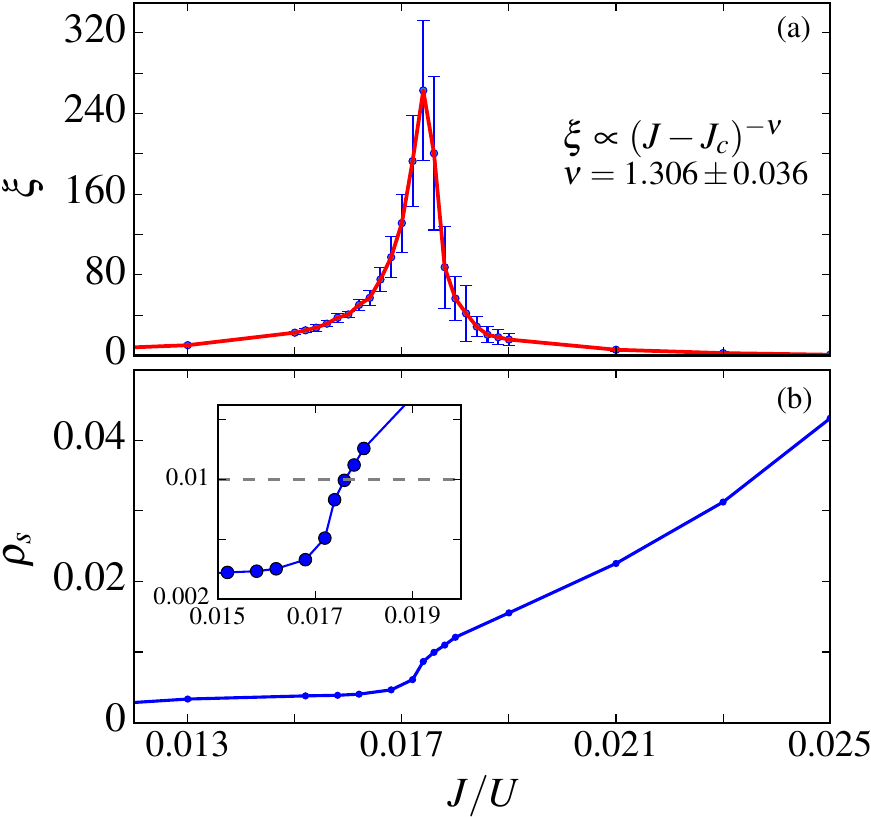}
    \caption{Plot of the correlation length $\xi$ and the superfluid stiffness
             $\rho_s$ as a function of $J/U$. The chemical potential is
             $\mu = 0.2U$ and disorder strength is $\Delta = 1.2U$ for these
             calculations. The results are obtained by averaging over 60
             disorder realizations. In (a), we observe a
             divergence of $\xi$, signaling the percolation transition from the
             BG to SF phase, with $J_c$ near $0.017U$. The critical exponent
             is $\nu = 1.306 \pm 0.036$. The standard deviation from the 
	     average value is shown by the blue error bar.
	     Panel (b) illustrates the stiffness $\rho_s$
             as a function of $J/U$. There is an increase in the 
	     stiffness near $J/U \approx 0.017$ as shown in the inset. 
	     For numerical calculations, the threshold value $0.01$ (shown in
	     gray dashed line) is considered to distinguish between the BG and 
	     SF phase.}
  \label{percol}
\end{figure}

In the Fig.\ref{percol} (a), the correlation length $\xi$ is shown as a 
function of $J/U$. As evident from the figure $\xi$  diverges near 
$J/U \approx 0.017$ and this signals a percolation transition from the 
BG phase to the SF phase. Using the relation in Eq. (\ref{corrlen_scalJ})
we can calculate the exponent $\nu$ which quantifies the divergence and 
obtain $\nu = 1.306 \pm 0.036$. This is in excellent agreement 
with the value $\nu = 4/3$ corresponding to the universality class of 2D 
random percolation model. Previous studies \cite{zuniga_15, niederle_13} 
have also computed the exponent $\nu$ and the values reported are in good 
agreement with our results. To study the BG-to-SF transition further, we plot 
$\rho_s$ as a function of $J/U$ in Fig. \ref{percol} (b). The calculation of 
$\rho_s$ are performed with a $100 \times 100$ system, as the computations 
with the twisted boundary conditions are compute intensive and require long 
execution times for a $1000 \times 1000$ system. As mentioned earlier, 
$\rho_s$ is small in the BG phase owing to the absence of a global phase 
coherence in the system.  It shows an increase as the system enters the SF 
domain.  Based on our previous work \cite{pal_19}, we consider 
$\rho_s \approx 10^{-2}$ as a threshold for distinguishing between the BG and 
SF phase. The plot shows an increase in $\rho_s$ at $J/U \approx 0.017$, and 
the threshold is crossed at $J = 0.0174U$. This is also the point where 
$\xi$ shows divergence. Thus, the identification of the BG-SF transition with 
the order parameter $\rho_s$ matches with the percolation analysis. Given the 
fine-grained approach of our method, the properties of the domain formation 
and their dynamical evolution can be analyzed using our method. This shall be 
addressed in our future works.


\subsection{Fractal dimension of hulls of SF clusters}
The percolating clusters near the percolation transition are self-similar
and can be characterized by a fractal dimension. In particular, the perimeter 
of the clusters close to the percolation transition point is similar to a 
random walk with many fractured edges and can be characterized by a fractal 
dimension $D$ which can be computed using the area-perimeter relation
\begin{equation}
  H \propto A^{D/2}
  \label{area-peri_reln},
\end{equation}
where $H$ is the perimeter or the hull and $A$ is the area of the cluster. 
In this subsection, we investigate the fractal properties and the compute 
the fractal dimension of the SF islands close to the BG-SF transition point.

\begin{figure}[ht]
    \centering
    \includegraphics[height = 4.3cm]{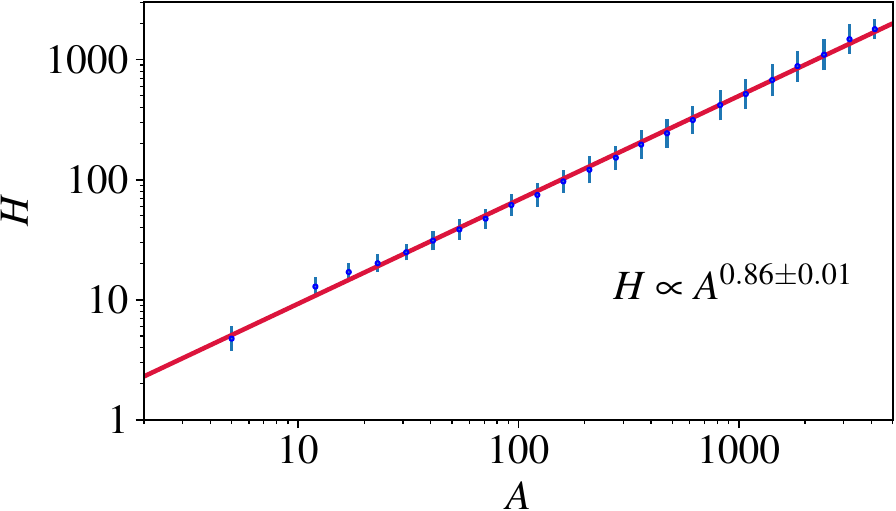}
    \caption{Power-law scaling of the hulls of the clusters with their area.
             The data points are represented by blue color, while the
             red line is the best fit to the data. The fractal dimension
             from this analysis is $D = 1.72$.}
    \label{hull_area_scal}
\end{figure}

For our studies, we consider $J = 0.017U$ for a $100\times 100$ system and 
consider $60$ disorder realizations. Once we identify the equilibrium 
configurations of each of these $60$ realizations and identify the SF islands, 
we use the boundary walk algorithm to compute the hulls of the clusters.
\begin{figure}[ht]
    \centering
    \includegraphics[height = 7cm]{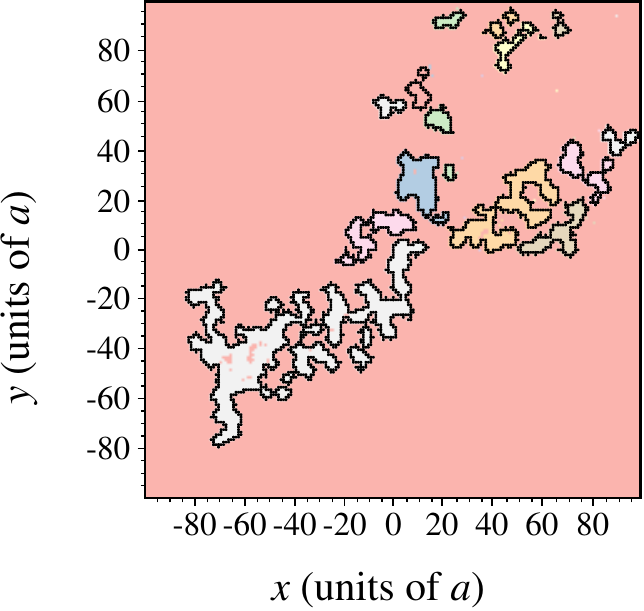}
    \caption{Charted boundary of the SF clusters shown in 
	     Fig.~\ref{percol_phi} (b) after unwrapping the system. The 
	     clusters in the original system are mapped onto the system with 
	     double dimensions. The SF clusters are identified with different
	     colors in the figure.}
    \label{bndry_1sample}
\end{figure}
As $J$ is close to criticality, there are clusters which lie across the 
edges along the $x$ and $y$ directions due to the periodic boundary conditions.
To simplify the analysis, we unwrapped the system and mapped the clusters onto
a larger system. This ensures that each cluster forms an island without 
touching the edges. Then, the starting site and the prior hop for the boundary 
walk can be accomplished without any ambiguity. As an illustration, we show 
the unwrapping of the order parameter profile corresponding to 
Fig.~\ref{percol_phi}(b) and the charted boundaries of these clusters in 
Fig.~\ref{bndry_1sample}. The hull of all the clusters in other disorder 
realizations are computed in same way. The variation in the hull of these 
clusters with the cluster area is shown in Fig.~\ref{hull_area_scal}. In this 
plot, the hulls of clusters with area in the bin $[A, A + \delta A]$ are 
averaged to give $H(A)$. We have chosen $\delta A = 0.3 A$ for our studies. 
The $\delta A \propto A$ assures uniform spacing in the log $A$ scale. 
We do a least-squares fit to the data using the area-perimeter scaling 
given in  Eq.~(\ref{area-peri_reln}) and the result is shown in 
Fig.~\ref{hull_area_scal}. 
The fractal dimension $D$ obtained from the results is $1.72 \pm 0.02$. This 
value is in good agreement with the predicted value of the fractal dimension 
$D = 1.75$ in the universality class of 2D random 
percolation \cite{voss_84,grossman_86,grossman_87,ziff_89}. Thus the analysis 
of the fractal dimension of the SF islands in the BG phase complements or 
confirms the earlier conclusion that the BG-SF transition belongs to the 
universality class of 2D random percolation. This is one example application
of the fine-grained analysis using the DC algorithm.


\section{Conclusion}
\label{conclusion}
 To summarize, we have developed an algorithm to identify clusters or domains 
and study their percolation properties on a 2D lattice. The algorithm 
exploits the structure of the linked-lists data type, and enumerates the 
clusters on the lattice with a single scan. Further, the locations of the 
sites constituting a cluster are stored, and this facilitates the application
of methods based on percolation theory. Our approach, with minor modifications
in the scanning process, can be adapted to lattices of higher dimensions.
Based on the results, we also present an algorithm to identify the boundary
of the cluster. Using the algorithm, we compute the number of domains
formed in the MI-SF quench dynamics and calculate the critical exponent.
The critical exponent so obtained is in excellent agreement with the result 
from the vortex density. Further, we also study the BG-SF transition in the 
DBHM. For this we calculate the correlation length of the SF clusters and 
observe a divergence near the BG-SF transition. In addition, we
 have computed the fractal dimension of the SF clusters near the BG-SF 
transition. The exponent $\nu$ associated with the divergence of correlation 
length and the fractal dimension $D$ suggests that the BG-SF transition belongs 
to the universality class of 2D random percolation.

\begin{acknowledgments}
The results presented in the paper are based on the computations
using Vikram-100, the 100TFLOP HPC Cluster at Physical Research Laboratory, 
Ahmedabad, India. D.A. acknowledges support from the Science and
Engineering Research Board, Department of Science and Technology, Government
of India, through Project No. CRG/2022/007099 and support from the UGC through
the SAP (DRS-II) Project No. F.530/18/DRS-II/2018(SAP-I), 
Department of Physics, Manipur University. The authors are grateful to 
Dr. Sukla Pal, Dr. Kuldeep Suthar, Dr. Rukmani Bai, and 
Dr. Soumik Bandyopadhyay for the fruitful discussions and their comments to 
improve the paper.
\end{acknowledgments}

\appendix


\section{Comparison with other algorithms}
\label{appendA}
We have compared the execution times for the domain identification required by
the HK, DC, and the recursive neighbor search algorithms. The comparison
is based on analysis of  randomly generated $40$ lattice configurations. In 
each member the sites are randomly labeled with $0$ and $-1$. As mentioned
earlier, the sites labeled $0$ are to be identified as a domain. For
comparison, for each member the execution time using the three algorithms are
noted. To remove the effects of fluctuation on account of system load and 
other factors, an average of the results is calculated from the $40$ members.
For uniformity and to avoid computational overheads, the FORTRAN implementation
of these algorithms are executed on a HP EliteDesk PC with Intel i7 processor.

The comparison between the execution times of the HK and the DC 
algorithms are shown in Fig.~\ref{time_comp}. For the 
HK algorithm we have used the FORTRAN implementation by Anders from 
the GitHub \cite{anders_2022}. The results indicate that the 
DC algorithm is marginally faster than the HK algorithm. The least-squares fit 
of the data gives $T_{\rm DC} = 9.58 {N_s}^{1.01}$ nanoseconds and 
$T_{\rm HK} = 16.05 {N_s}^{1.00}$ nanoseconds, where 
$T_{\rm DC} (T_{\rm HK})$ denote the execution time required in the DC (HK)
algorithm. This suggests that both these methods have 
number of floating point operations proportional to the number of sites. 
However, it is to be noted that the execution time of the DC algorithm also 
includes the construction of the linked lists, and these lists can be utilized 
for further studies. 
The main panel in Fig.~\ref{time_comp} includes time required for the first
scan and the following postprocessing steps. In the DC algorithm, the 
postprocessing step is to reassign the cluster labels in ascending order, as 
the cluster labels after the first scan are not in a sequence. 
This step is thus to make the labels systematic which helps
for percolation studies. However, post the first scan, the clusters are 
identified with unique labels thus completing the domain identification process.
On the contrary, the second scan in the HK algorithm is necessary to label the
clusters with unique labels.
This further adds to the advantages of the DC algorithm. 

The domain mergers are the reason for the postprocessing required in the DC 
and HK algorithms. However, there are no domain mergers in the recursion 
method, and hence there is no postprocessing of labels. Therefore we have 
compared the run time of the algorithms until the first scan, avoiding 
the postprocessing of the data, and the results are shown in the inset in 
Fig.~\ref{time_comp}. We have illustrated the comparison for a small range of 
lattice sizes for the readability of the figure. We notice that the DC 
algorithm 
requires extra time, around $4 \%$ more relative to the HK algorithm. This is 
due to the construction of the linked-lists. This also suggests that the 
postprocessing in the DC algorithm requires lesser time than that in the HK 
algorithm. This is to be expected, as in the DC algorithm, the 
postprocessing involves transfer of the linked-lists from the old to new 
cluster labels, without involving a second scan. However, for the HK algorithm,
the postprocessing is the secondary lattice scan.
It is also evident that the recursive neighbor search is 
slower than the DC and the HK algorithm, which is another disadvantage of this 
method. 

\section{System Unwrapping}
\label{appendB}
We describe the unwrapping procedure employed on the system required to compute
the distances within the system. The aim is to map the system of size 
$L_x \times L_y$ with periodic boundary conditions onto a larger system of size
$2L_x \times 2L_y$ with open boundary condition. To begin with, we first 
identify the clusters on the boundary of the system. For illustration,
consider a cluster that straddles across the left and right edge of the system.
Due to the PBC, the left and right chunks are a part of same cluster. The chunk
of this cluster connected with the right edge of the system needs a $x$ shift. 
The $x$ shift implies that the chunk is translated along the negative $x$ 
direction
by distance $L_x$.  After this step, the two chunks appear side-by-side and the
distance between any two points is unique. Similarly a $y$ shift is assigned to
a chunk of the cluster which straddles along bottom and top edge. In some 
cases, 
a cluster may straddle across both directions. In this case, certain chunks may
require both $x$ shift and $y$ shift. As discussed in the main text, 
the unwrapping of the domain configuration of Fig.~\ref{percol_phi}(b) is shown
in Fig.~\ref{bndry_1sample}. After these steps are performed, we compute 
the geometrical properties of the clusters. Here we mention that these steps 
are not valid for percolating clusters as they may be infinite in length, and 
would not accommodate in a new system of finite size. But as stated in the main
text, the percolation analysis of the system generally ignores the contribution
from the percolating clusters, thereby shifting of the finite sized 
nonpercolating clusters suffices to perform the percolation studies.


\bibliography{dom_cnt}{}

\end{document}